# ASL Champ!: A Virtual Reality Game with Deep-Learning Driven Sign Recognition

Md Shahinur Alam
VL2 Center, Gallaudet University, Washington, DC, USA;
md.shahinur.alam@gallaudet.edu

Jason Lamberton
VL2 Center, Gallaudet University, Washington, DC, USA;
jason.lamberton@gallaudet.edu

Jianye Wang
VL2 Center, Gallaudet University, Washington, DC, USA; Jianye.wang@gallaudet.edu

Carly Leannah
VL2 Center, Gallaudet University, Washington, DC, USA; carly.leannah@gallaudet.edu

Sarah Miller
VL2 Center, Gallaudet University, Washington, DC, USA; sarah.miller@gallaudet.edu

Joseph Palagano
VL2 Center, Gallaudet University, Washington, DC, USA;
joseph.palagano@gallaudet.edu

Myles de Bastion
CymaSpace, Portland, OR, USA; myles@cymaspace.org

Heather L. Smith
OneMonkey Studios LLC, Rochester, NY, USA; hszvfx@gmail.com

Melissa Malzkuhn
VL2 Center, Gallaudet University, Washington, DC, USA;
melissa.malzkuhn@gallaudet.edu

Lorna C. Quandt
VL2 center, Gallaudet University, Washington, DC, USA; lorna.quandt@gallaudet.edu

*Corresponding author. lorna.quandt@gallaudet.edu, 434-294-6458.
Gallaudet University SLCC 1220
800 Florida Ave NE
Washington, DC USA.

We developed an American Sign Language (ASL) learning platform in a Virtual Reality (VR) environment to facilitate immersive interaction and real-time feedback for ASL learners. We describe the first game to use an interactive teaching style in which users learn from a fluent signing avatar and the first implementation of ASL sign recognition using deep learning within the VR environment. Advanced motion-capture technology powers an expressive ASL teaching avatar within an immersive three-dimensional environment. The teacher demonstrates an ASL sign for an object, prompting the user to copy the sign. Upon the user's signing, a third-party plugin executes the sign recognition process alongside a deep learning model. Depending on the accuracy of a user's sign production, the avatar repeats the sign or introduces a new one. We gathered a 3D VR ASL dataset from fifteen diverse participants to power the sign recognition model. The proposed deep learning model's training, validation, and test accuracy are 90.12%, 89.37%, and 86.66%, respectively. The functional prototype can teach sign language vocabulary and be successfully adapted as an interactive ASL learning platform in VR.

Keywords: Virtual reality, deep learning, american sign language, interactive learning, VR learning, avatar interaction,

## 1 INTRODUCTION

Sign language plays a vital role in the lives of numerous individuals. These are natural, full languages developed within the deaf or hard-of-hearing communities. Every sign language employs a distinct set of manual signs and distinct body movements. Over 5% (430 million) of the world's population has some form of hearing loss, which is projected to increase to 2.5 billion by 2050 (*Deafness and Hearing Loss,* n.d.). Signed languages are natural languages, which means they exhibit unique characteristics stemming from the surrounding culture, ethnicities, and geographical regions where they evolve. Sign language proficiency is rare among the global hearing population, thus necessitating interpreters for many daily interactions. As emerging technologies advance, using these technologies to create new avenues for sign language learning is promising. Furthermore, recent developments in immersive technologies, such as virtual reality (VR), offer exciting educational prospects, including the potential for immersive learning and interaction with signed languages within VR environments. Accurate recognition of users' signing is paramount for effectively teaching signed languages in VR settings (Alam et al., 2023; L. Quandt, 2020), as previous ASL learning systems in VR without feedback failed to provide meaningful interactive experiences for users. Teaching and learning ASL with technology is still of limited use, given the complexity of ASL and current technology capability. While numerous resources for learning sign language exist, none offer the feedback that real teachers provide in a physical classroom setting. It's best for ASL learners to learn from teachers who are native ASL users and give real-time feedback(Tseng et al., 2013). Thus, creating an autonomous learning system incorporating feedback is imperative for creating an engaging educational experience.

Recent studies on ASL recognition algorithms typically use deep learning (DL) algorithms. Surveys indicate that DL-based algorithms exhibit notable improvements in accuracy over classical/machine learning algorithms (Barve et al., 2021; Fatmi et al., 2019). Given that VR devices are equipped with low-computational-power embedded microprocessors, designing a lightweight DL network is necessary (Meske et al., 2022). Certain applications necessitate both object detection and the identification of the user's hands, requiring a complex DL network (Kang et al., 2020; Mao, 2022), mostly a convolutional neural network. The field of ASL recognition is experiencing rapid growth (Hays et al., 2013; Pugeault & Bowden, 2011; L. Quandt, 2020; Shao et al., 2020; Sharma & Kumar, 2021; Vaitkevičius et al., 2019). The commonly used approach for ASL recognition involves two-dimensional (2D) cameras or wearable devices, but these methods are less efficient and often impractical in real-life scenarios (Thongtawee et al., 2018; Wen et al., 2021). ASL communication encompasses hand gestures, facial expressions, body postures, spatial cues, and dynamic movements, making combining all aspects into 2D systems challenging. Research has shown that wearable ASL recognition devices can be problematic, and they have not garnered significant interest from the signing communities (Hill, 2020; Zhou et al., n.d.). In some cases, VR devices exhibit reasonably good recognition outcomes (Schioppo et al., 2019; Vaitkevičius et al., 2019). However, many existing efforts rely heavily on the Leap Motion and do not constitute full-fledged VR systems.

Consequently, developing a standalone ASL recognition system within a VR environment remains an unresolved challenge. A critical part of this system is incorporating feedback to inform the users whether their sign productions are correct. This feedback relies on capturing and analyzing users' sign expressions using the built-in camera sensors of the VR device. To address the limitations mentioned earlier, we have developed a VR ASL recognition system trained on a highly diverse set of signed inputs. We designed an ASL learning game that runs on standalone VR headsets, incorporating ASL recognition, to create an interactive learning and feedback system for adults new to ASL. In this context, we focus on developing a simple DL network that can efficiently be implemented within a VR environment. As part of the larger work (Alam et al., 2023; L. Quandt, 2020; L. C. Quandt, Lamberton, et al., 2022), we aim to teach ASL using a virtual reality game-like environment. In this virtual setting, users will immerse themselves and learn from signing avatars created through motion capture recordings. The main contributions of this paper are as follows:

- We developed a fully functional ASL learning environment in VR called *ASL Champ!* In this study, we designed an immersive, naturalistic environment to teach introductory ASL vocabulary.

- We built a training dataset from the ASL sign data of fifteen participants, each producing nine ASL signs repeated ten times. These signs serve as the training data for the recognition system.

- We introduce a deep learning model designed for ASL sign recognition. This model offers reasonable recognition accuracy.

- The findings of the user study reveal significant insights into the impact of ASL learning in VR and provide valuable guidance for future research directions in this domain.

## 2 BACKGROUND AND CHALLENGES

Advancements in technology have led to the development of interactive educational environments, including immersive 3D learning platforms referred to as virtual reality (Solomon et al., 2019; Wang et al., 2021). VR is increasingly acknowledged as a valuable tool for education. Traditionally, VR is linked to entertainment and has had some emerging impact on the education system (Kamińska et al., 2019), manufacturing (Choi et al., 2015), healthcare (Javaid & Haleem, 2020), etc. Educational institutions increasingly use VR to teach complex abstract concepts and expand teaching beyond regional boundaries (Wang et al., 2021). ASL learning in augmented reality (AR) and VR is limited. An AR-based ASL recognition has been implemented to a small extent (Soogund & Joseph, 2019). SignAR aids deaf and hard-of-hearing students in learning English and sign language through word-to-animation mapping. It improves academic performance and opportunities by improving communication between hearing and deaf people. However, avatar-based VR interaction can be vital in teaching ASL in a game-like environment

(L. Quandt, 2020). Signing avatars offer valuable, accessible language resources for the deaf community. Creating avatars that sign using high-fidelity motion capture of native signers results in more naturalistic avatars (L. C. Quandt, Willis, et al., 2022). Although augmented reality has been used to teach sign language, immersive VR experiences for learning sign language are currently unavailable in the public domain due to limitations, such as implementation difficulties and the unavailability of dynamic content. The environment in which we acquire a new language, especially for adult second language learners, can profoundly impact important language learning metrics such as proficiency, emotional processing (i.e., motivation), and memory recall performance (Chun et al., 2016). Immersive and embodied contexts of learning have been shown to support these metrics.  Content generation in VR is a tedious and complex job. Recently, we developed an ASL learning platform in VR, focusing on ASL numeric digits (Alam et al., 2023). However, ASL is a fully developed language, and implementing only numeric signs is insufficient to create a usable learning platform. Additionally, the digits represent a very simple subset of the ASL lexicon and are static. Dynamic, moving signs require a more complex algorithm for recognition. As motion is a foundational parameter in sign language linguistics, a subtle change in the kinesthetic production of a sign can convey an entirely different meaning. For instance, the signs for CHOCOLATE, CHURCH, and COMPUTER in ASL are minimal pairs, which means they all use similar handshapes and locations but differ only in their motion trajectories.

Developing an efficient algorithm to handle the complexity associated with dynamic signs posed a substantial challenge in our research, and the following sections will dive into the details of our methodology and findings in overcoming these obstacles. The primary challenges to developing effective ASL instruction in VR are the lack of VR datasets, implementation difficulties, and the underrepresentation of deaf/native signers in development. Our approach involves the development of a VR environment simulating a naturalistic environment, wherein a list of basic signs will be taught. This work aims to encompass ASL's dynamic and nuanced nature. The work we present here attempts to mitigate the following three challenges.

## 2.1 Lack of VR Datasets

A notable challenge in the field of ASL recognition is the limited availability of sign language datasets, particularly in the context of VR environments. While some researchers have begun to address this issue, datasets are still scarce. For instance, Pugeault et al. compiled a dataset comprising 131,000 ASL alphabet samples using Kinect sensors, OpenNI, and the NITE framework (Pugeault & Bowden, 2011). This dataset has certain limitations, such as incomplete coverage of the ASL alphabet, encompassing only 24 of the 26 characters and focusing solely on static signs. Most of the current datasets primarily consist of 2D data such as – WLASL (Li et al., 2020), RWTH-Phoenix-2014T (Camgoz et al., 2018), MS-ASL (Joze & Koller, 2018), etc. However, How2Sign is a comprehensive ASL dataset featuring over 80 hours of sign language videos and related modalities such as speech, English transcripts, and depth information. Additionally, a three-hour subset was recorded in the Panoptic studio for precise 3D pose estimation. While this dataset is well-suited for a 3D environment, it presents challenges when integrating into VR due to differences in VR coordinates. Additionally, incorporating RGB-D data into VR can be computationally demanding, especially on systems with limited computational power.

## 2.2 Implementation Difficulties

Unlike other gesture-based interaction systems (Alam et al., 2019, 2021), real-time ASL identification is a difficult task that requires carefully recording and analyzing subtle aspects such as hand, fingers, body, and gaze movements. These elements blend together simultaneously to make the intricate pattern of sign language communication. One of the most significant issues is the possibility of occlusion, which occurs when a signer's hands or arms hide other elements of sign language production. Furthermore, differences in the distance between the signer and the capturing device might cause issues since the scale and perspective of the signs may vary. Furthermore, lighting conditions might change significantly, impacting the quality and clarity of the visual data recorded. Another issue is the possibility of color ambiguity within signs, which can occur due to variations in skin tones or clothes. To adequately address these problems, an ASL recognition system must be robust, which means it must be flexible and resilient enough to appropriately identify the subtle variances that occur naturally in sign production. When implementing ASL recognition in a VR context, achieving this degree of accuracy becomes increasingly difficult. VR usually has limited processing capability, making high-end deep learning models commonly utilized in ASL identification systems difficult to run (Ibrahim et al., 2019).

## 2.3 Underrepresentation of native/fluent ASL signers

Many existing ASL datasets were compiled from hearing people with little or no ASL proficiency, which may introduce significant errors in their sign production (Schlehofer & Tyler, 2016). Even individuals who have undergone years of sign language instruction fail to produce proper signs, frequently making errors in sign movement, placement, and orientation (Barve et al., 2021). There is a risk of producing homogeneous datasets when training ASL recognition models with data from inexperienced signers. This homogeneity can be seen in signs routinely created in the same location or orientation for each sign occurrence. In the real world, ASL is used by people of various competency levels, complete with individual differences in sign production, including handedness, such as left and right-handed individuals. These differences include spatial characteristics, pacing, or speed. This intrinsic natural variability is one of the primary reasons why the accuracy of most ASL recognition models falters when used in practical, real-world contexts. To address this constraint, it is critical to create comprehensive and varied datasets that cover a broad spectrum of sign output by people with varying degrees of competence.

We used a variable signed input dataset to train a VR-based ASL recognition system. Rather than placing constraints on signers, we encouraged them to spontaneously produce the sign they use for the English word equivalent (e.g., produce the ASL sign MILK for the English word "milk"), replicating the wide range of signs observed in the actual world. By taking this approach, we aim to address the difficulties associated with ASL recognition's real-world applications.

## 3  DATASET CREATION

### 3.1  ASL data collection

The ASL champ game has two distinct user interfaces (UI's)- data collection and ASL learning environment (coffee shop). Labeling data is imperative for robust algorithmic training because the feedback system uses a supervised DL algorithm. An intuitive data collection UI has been developed to create an accurate and diverse ASL VR dataset. Figure 1 depicts the UI we developed for gathering ASL sign data in VR. Within this UI, users have the freedom to integrate their data into custom datasets. Although the system offers a default database, users possess complete control to tailor and rename it, allowing them to create new databases. For instance, they can select the sign name from the dropdown under "Sign name" to label and categorize their signs. We included all sign names in this dropdown list. Users are free to adjust the duration of the sign they record, unlike in previous research projects where fixed sign durations were typically employed despite variations in natural signing tempo among individuals. This user-driven flexibility was purposefully designed to accommodate a greater spectrum of signers.

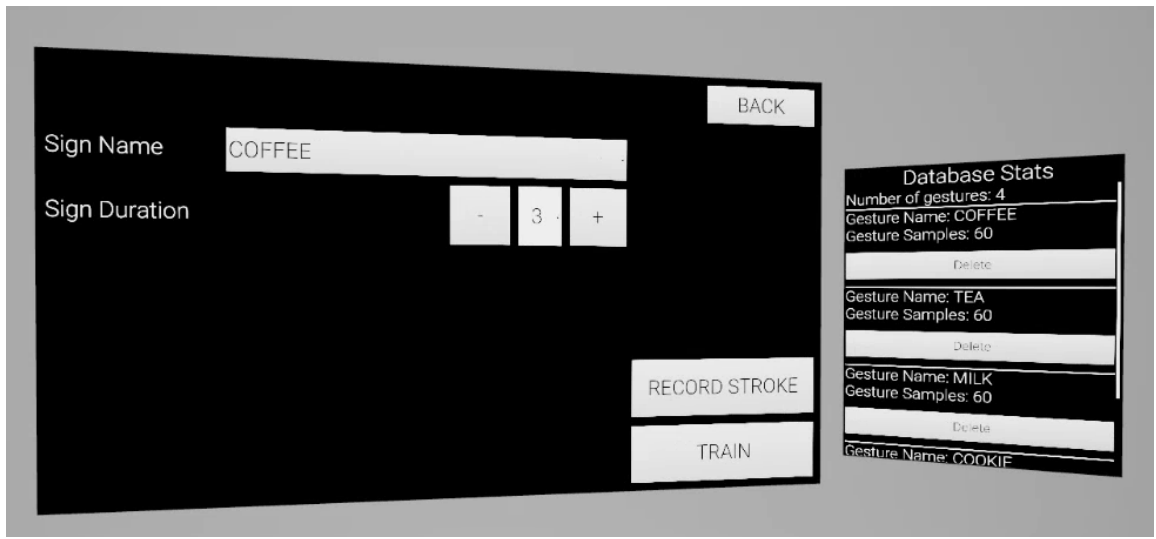

Figure 1. The data collection UI contains dropdown options for sign selection, including record and train buttons. Users can interact with this UI to produce signs to train and store ASL data.

Initially, the "RECORD STROKE" and "TRAIN" buttons are invisible. When a user initiates the system by selecting the sign name, the "RECORD STROKE" button becomes visible. Similarly, the "TRAIN" button becomes visible after a successful sign record. The participant can see the list of signs and the number of signs in the right window. They can delete specific signs if needed. Our previous experiments showed that the user is sometimes confused about which button to press first (Alam et al., 2023). Our new step-by-step process helped solve that issue. When the participant initiates the recording process, the system commences the real-time tracking of hand and joint movements for the specified duration. Once this phase is complete, participants are prompted to tap the "TRAIN" button. This action serves a dual purpose: it saves the recorded gesture to local storage for future reference and initiates the training process for the MiVRy (*MiVRy 3D Gesture*

*Recognition in Code Plugins - UE Marketplace*, n.d.) model. During the training phase, the system uses the obtained data to train the network to detect and recognize the sign.

3.2 **Dataset parameters**

Our data collection encompassed ASL signs, specifically focusing on capturing both hand information and hand-joint data. On each hand, we recorded data from 25 joints. Each joint entry had three crucial data points: bone location, rotation, and hand rotation information (local), all precisely collected using the Oculus API. Typically, the duration assigned to most gestures was 3 seconds, storing 217 data points per sign. In the location data of each joint, the local coordinates in the x, y, and z dimensions were collected. Additionally, pitch, yaw, and roll were recorded for rotational data.

We opted to design a coffee shop environment and visited a nearby coffee shop where ASL is commonly used to ensure that we selected signs used in real-world signing interactions (*Signing Starbucks Ranked "Visit-Worthy" | University Communications | Gallaudet University*, n.d.). We selected nine ASL signs for the ASL Champ! prototype: COFFEE, TEA, MILK, WHIPPED CREAM, MUFFIN, COOKIE, CUP, STRAW, and MONEY.

4 **METHODS**

The following section will describe the methodologies and approaches for ASL recognition in a VR environment. We discuss technical challenges like capturing hand movements while considering VR's processing limits. We will describe the creation of VR ASL recognition systems, emphasizing hardware, software, and user interface options.

## 4.1 Developing the 3D environment

To design an effective ASL learning game, we developed an immersive and engaging 3D coffee shop. This semi-realistic environment was created to provide a location for interaction with the signing avatar and allow a space for producing the signs and receiving real-time feedback. We used a multi-step procedure to do this. First, we created sophisticated and detailed 3D models, textures, and materials using 3D modeling software, notably Autodesk Maya. This included creating a 3D building, props, and a teacher character for the coffee shop environment. These 3D components were designed to simulate the mood and appearance of a real coffee shop.

Following that, we integrated these components into the Unreal Engine environment, elevating the overall user experience by combining the realism of our 3D graphics with the immersive potential of the Oculus Quest. The coffee shop scene was improved using standard lighting techniques. This attention to detail was intended to make the scene as pleasant as possible, allowing participants using the Oculus Quest headset to have a more immersive and engaging learning experience.

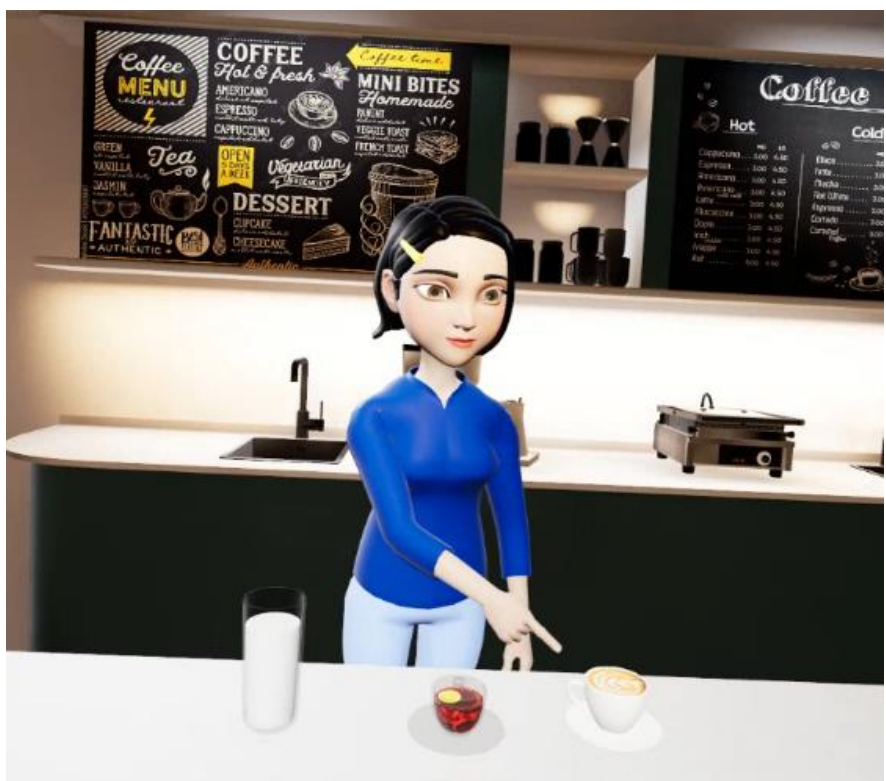

Figure 2. The 3D coffee shop environment. The avatar points to indicate the sign referent before producing the sign for COFFEE.

Figure 2 illustrates our VR coffee shop environment, which serves as our initial lesson. Three objects sit on a counter at the start of this lesson: milk, coffee, and tea. The avatar demonstrates the sign for each item and then loops back with one repetition, so the sign for each item is taught twice. This aims to make learners feel like they are interacting with a person teaching them signs in a coffee shop. In an ASL class, teachers often repeat new signs so students have time to learn how to produce the movements for a particular sign. This practice aligns with work demonstrating that hands-on experience with new material results in greater "embodied learning" or sensorimotor representation of the practiced content (Cannon et al., 2014; Kontra et al., 2015; L. C. Quandt & Marshall, 2014).

## 4.2 **Motion capture**

The teacher avatar was created through motion capture technology, starting with real individuals as the source of motion. The motion data is captured, post-processed, and polished to ensure accuracy and usability. This comprehensive approach captures the signer's body, hand, and finger movements, making it a robust foundation for creating avatars who sign well. During motion capture sessions, we used a state-of-the-art Vicon system with 18 high-resolution cameras (8 T160 cameras, 10 Vero cameras) and 73 markers carefully placed on the signer's fingers, hands, and body. We used Vicon Shogun 1.7, a tool that significantly enhanced the quality of motion capture, particularly in capturing body, hand, and finger movements with high fidelity. The full system is shown in Figure 3.

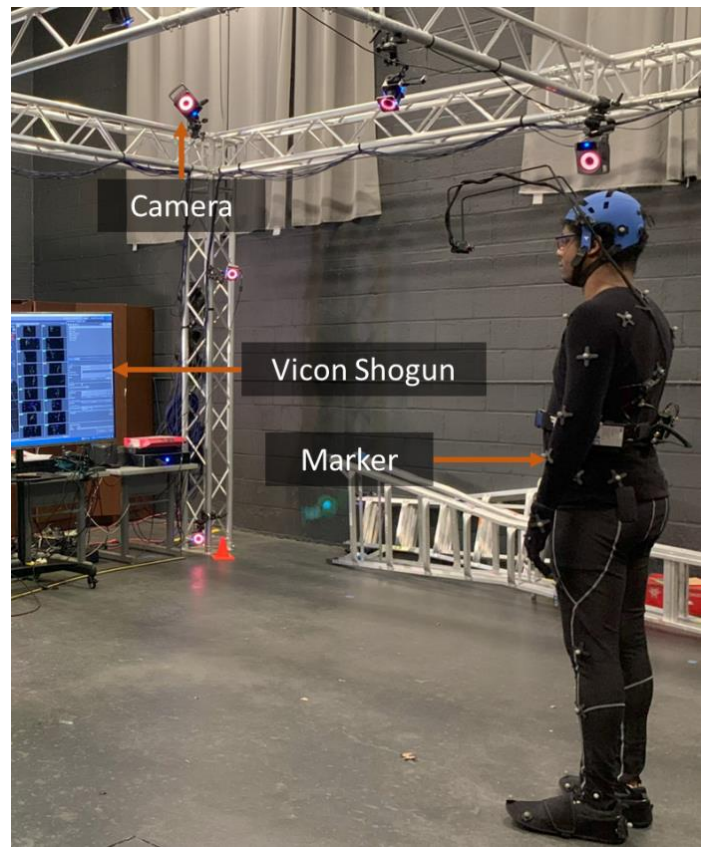

Figure 3. The motion capture system The model is prepared to produce the signed content by attaching markers on the velcro body suit using the recommended configuration.

We conducted the motion capture sessions with a native deaf signer as the model. This choice ensured that the captured data accurately represented natural sign language hand and body movements. This approach to avatar creation underscores our commitment to delivering an engaging and immersive signing experience for users.

## 4.3 Sign Detection

The critical component of the feedback system in our learning platform is sign recognition, where we have employed two distinct methods: a third-party tool and our proprietary deep-learning model. Details of these approaches are outlined below.

### 4.3.1 *Third-party tool*

We've integrated the MiVRy Unreal Engine plugin into our platform as a third-party tool for sign recognition capabilities. This powerful tool serves as a bridge, enabling even non-programmers to train and recognize signs and gestures effectively. MiVRy leverages advanced artificial intelligence to identify hand gestures within VR and various applications, offering flexibility during the initial prototype stages. With MiVRy, users can record their own hand/finger gestures or invent new ones on the fly. The AI learns to recognize these motions and offers detailed feedback, including the identified gesture and its unique parameters within the virtual environment, such as location, direction, and scale. While MiVRy is useful, its recognition accuracy did not meet the expectations (Alam et al., 2023). To address this, we created our own deep-learning model.

### 4.3.2 *Deep learning model*

The proposed neural network design combines the characteristics of convolutional neural networks (CNN) and long short-term memory (LSTM) networks to improve feature extraction and sequence recognition tasks.

The core building block of a CNN is the convolutional layer. It involves applying convolution operations to the input data. The convolution operation is defined as:

$$(f * g)(t) = \sum_{a=1}^{m} f(a). g(t - a)$$

here, $f, g, t$, and $m$ represent the input data, filter position in the output, and the size of the filter, respectively. '$*$' denotes the convolution operation. Primarily, it extracts features by sliding filters over input data. This operation introduces translation invariance, allowing the network to recognize patterns irrespective of their position. Parameter sharing reduces the number of learnable parameters, ensuring efficiency and generalization. Overall, the convolution operation empowers CNNs to learn and represent complex spatial features efficiently, making them highly effective for tasks like image recognition.

An LSTM network is a recurrent neural network (RNN) designed to address the vanishing gradient problem and capture long-range dependencies in sequential data. LSTMs use memory blocks to store and control the flow of information over time. A basic LSTM memory block is shown in Figure 4. The main block components are input, hidden gate, forget gate, cell state, and output gate.

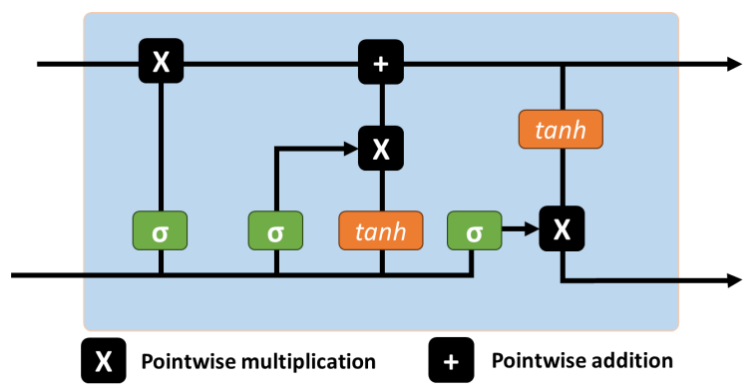

Figure 4. An LSTM memory cell.

The calculations within an LSTM memory block are governed by the following equations:

$$i_t = \sigma(W_{ii}.x_t + b_{ii} + W_{hi}.h_{t-1} + b_{hi})$$

$$f_t = \sigma(W_{if}.x_t + b_{if} + W_{hf}.h_{t-1} + b_{hf})$$

$$g_t = tanh(W_{ig}.x_t + b_{ig} + W_{hg}.h_{t-1} + b_{hg})$$

$$o_t = \sigma(W_{io}.x_t + b_{io} + W_{ho}.h_{t-1} + b_{ho})$$

$$C_t = f_t.C_{t-1} + i_t.g_t$$

$$h_t = o_t.tanh(C_t)$$

here, $x_t$, $h_{t-1}$, $W$, and $\sigma$ is the input at time step $t$, the hidden state from the previous time step, weight matrices, bias vectors, and sigmoid activation function, respectively. These equations describe how the input, forget, and output gates control the flow of information into and out of the cell state, allowing LSTMs to selectively remember or forget information over long sequences. The tanh function is used to regulate the values in the cell state.

After the convolution and LSTM operations, an activation function is applied element-wise to introduce non-linearity. The tanh function is frequently used here:

$$f(x) = \frac{e^x - e^{-x}}{e^x + e^{-x}} \qquad (1)$$

here, $e$ and $x$ is the mathematical constant and input to the function, respectively. A regular tanh function is shown in equation (1), x is any given value. It is defined as the quotient of the difference between the exponential of x and the exponential of negative x, divided by their sum, and it maps real numbers to a range between -1 and 1, exhibiting an S-shaped curve. It is notable for its symmetrical properties around the origin, being zero-centered, which means it maintains an average output value close to zero, making it useful for data normalization. In our dataset, the presence of both negative and positive values hold substantial importance, as they each contribute significantly to our model. This is precisely why we have utilized the tanh activation function in both our convolutional and dense layers.

A dense layer, also known as a fully connected layer (FCL) in a neural network, involves each neuron being connected to every neuron in the preceding layer, forming a comprehensive connectivity pattern. This layer, often found in the final stages of a neural network, introduces non-linearity through activation functions and determines the network's output dimension. Employed in classification tasks, the FCL contributes significantly to learnable parameters, enhancing the model's ability to capture complex relationships.

Figure 5 depicts the whole network diagram, including components such as input layers, convolutional layers, pooling layers, LSTM units, and dense layers. The input data is sequential, with the dimensions of the input layer set to accommodate the maximum lengths of trajectory sequences, which is 651 in this context. The first convolutional layer utilizes a filter size of 512, followed by a max-pooling layer designed to downsample the CNN network. This downsampling not only reduces computational complexity but also contributes to cost efficiency. A filter size of 256 is employed for the second convolutional layer, which, like the first, includes a max-pooling layer. The primary role of the CNN layers is to generate essential features crucial for subsequent processing.

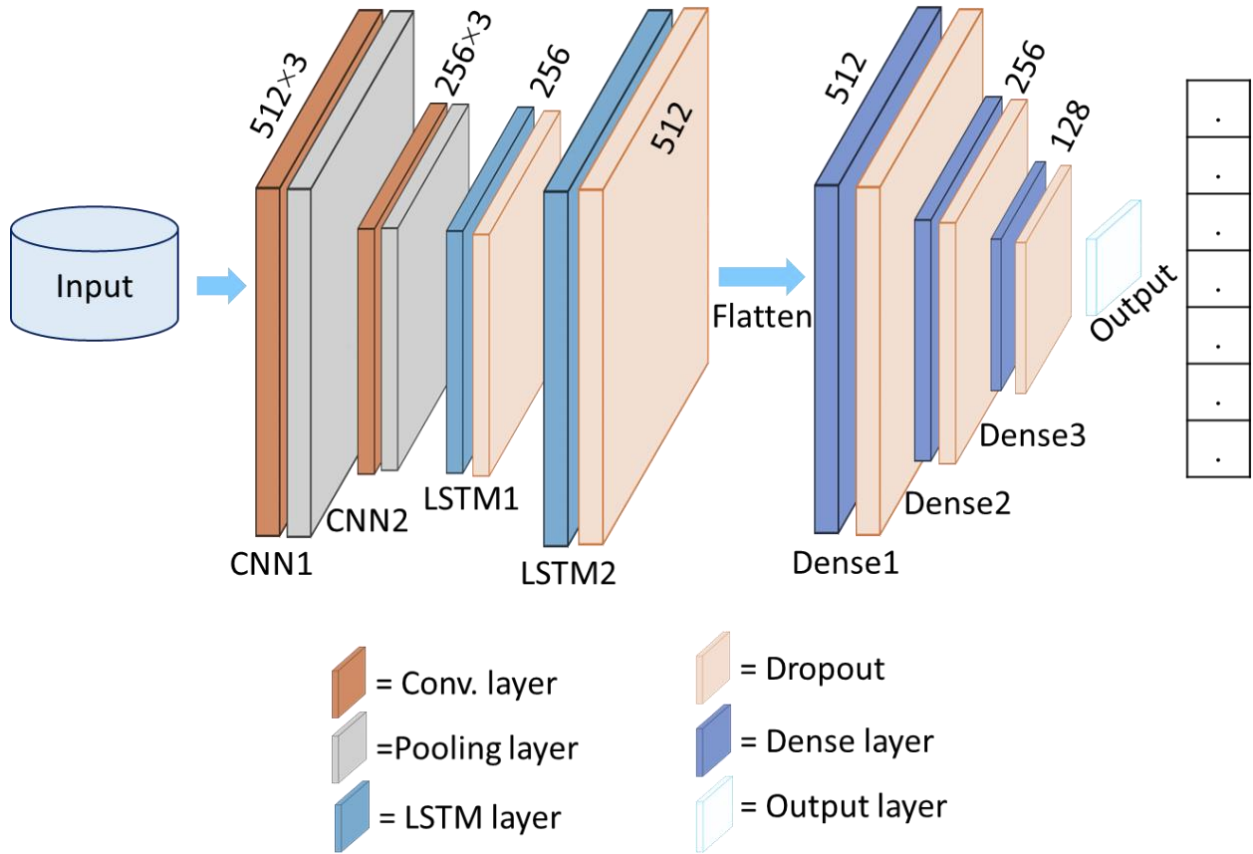

Figure 5. Current deep learning model. The model consisted of input, CNN, LSTM, dense, and output layers.

The network's LSTM component is divided into two layers, with the first layer containing 512 units and the second containing 256 units. Operations within the LSTM layer include sigmoid and hyperbolic tangent (tanh) activations, pointwise addition, and multiplications, allowing the network to capture detailed sequential patterns and relationships in the input. The data is flattened before proceeding to the dense layers to prepare it for further processing. The first dense layer has 512 neurons, followed by 256 neurons in the second dense layer and 128 neurons in the third dense layer. These dense layers are crucial in consolidating and refining the extracted features, improving the network's capacity to recognize complicated sequences and patterns in the input data. A dropout rate of 0.6 is applied to every dense layer to mitigate overfitting. This precautionary measure prevents the neural network from excessively relying on individual neurons, promoting a more resilient and broadly applicable model. This approach contributes to improved model performance and generalization, making the network better equipped to handle unseen data and reducing the risk of overfitting.

Two activation functions are used in this model- tanh (eq. 1), and SoftMax (eq. 2). The hyperbolic tangent function, commonly denoted as tanh(x) or more simply tanh, Tanh is frequently used as an activation function in neural networks, particularly in hidden layers of recurrent neural networks and LSTMs, where it introduces non-linearity and helps alleviate vanishing gradient problems.

$$\sigma(z)_j = \frac{e^{z_j}}{\sum_{k=1}^{K} e^{z_j}} \qquad (2)$$

The SoftMax function takes an input vector, often referred to as logits, and transforms it into a probability distribution where each element in the output vector represents the probability of belonging to a particular class. It does this by exponentiating each input vector element and normalizing the results. Equation (2) represents a SoftMax function where $\sigma(z)$ is the probability that the input belongs to class $i$, $Z_i$ is the $i$-th element of the input vector, $K$ is the total number of classes. The denominator is the sum of the exponentials of all the elements in the input vector, ensuring that the probabilities sum up to 1. The SoftMax function essentially amplifies the largest values in the input vector while suppressing smaller ones, resulting in a clear distinction between the probabilities assigned to different classes. We applied the SoftMax activation function in the output layer to generate only one output.

In this deep-learning model, we employ the Adam optimizer (Kingma & Ba, 2014) with categorical cross-entropy as our chosen loss function. These optimization and loss components were chosen to improve the network's training and performance. The Adam optimizer enables efficient convergence throughout training by dynamically changing the learning rates for each parameter. This adaptability is especially useful when dealing with complicated, high-dimensional data. It allows the network to fine-tune its parameters successfully.

## 5  SIGN DATA COLLECTION SETUP

Figure 6 depicts our data collection setup, including user interface. Signers wore an Oculus Quest 2 headset and produced ASL signs in the space before them. The UI was presented on both the computer monitor and the Oculus Quest 2 headset (running software version 44.0.0.169.455) at the same time. In this figure, the participant is signing TEA.

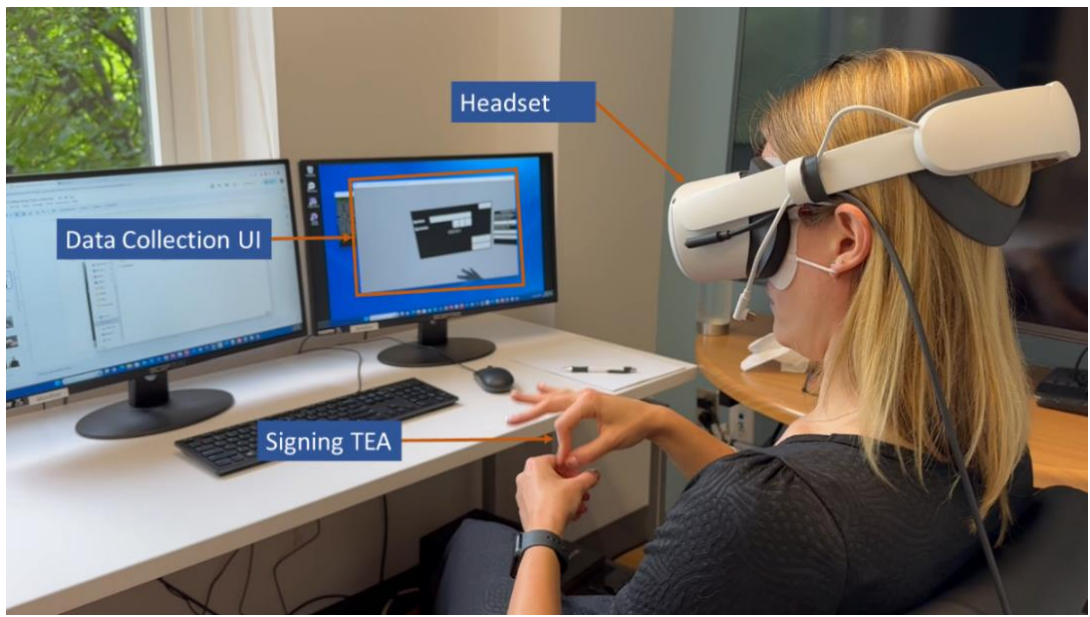

Figure 6. Sign collection setup: A signer in front of a computer wears a VR headset while signing "TEA.".

The UI was designed using Unreal Engine 4.27, incorporating the MiVRy plugin v2.5 for sign detection. To train our deep learning model, we used PyCharm and Keras on top of the TensorFlow API. The system operated on a Windows 11 Pro 64-bit operating system with 96GB of memory and was powered by an Intel Core i9 processor clocked at 3.50GHz.

One notable feature of our system is its flexibility in accommodating left- and right-handed signers. We encouraged signers to express themselves naturally, permitting variations in palm orientation and signing location. This adaptability is essential to capture the rich diversity of sign production styles observed in real-world ASL interactions. Each participant was tasked with signing each of the nine signs 20 times, leading to a substantial dataset with 180 signs from each participant. A total of 2,700 ASL signs were collected from fifteen signers. To ensure a diverse and comprehensive dataset for training our system, we enlisted the participation of 15 individuals, comprising four men and 11 women. The participants, aged 21 to 47, came from various linguistic backgrounds. Notably, eight of them had been exposed to ASL from birth, ten were deaf, and 85% were familiar with VR. This heterogeneity in language proficiency and VR exposure aimed to equip our system with the capability to recognize an extensive array of signing styles, thereby enhancing its real-world applicability.

## 6 PROTOTYPE RESULTS

In the context of the ASL Champ! game, the teacher avatar provides real-time feedback about whether the produced sign is correct. When users enter ASL Champ!, they are greeted by the teacher signing "Welcome to the coffee shop. Now I will show you some signs. Ready?" Since the primary goal of this work is to teach ASL to new signers, we included English captions for the welcome message. After the welcome message, the avatar demonstrates each sign twice, pointing to the corresponding object depiction on the counter in front of them. Then, the avatar prompts users to try producing the sign themselves. The avatar teaches a series of three signs at a time (e.g., in Fig 7, MILK, TEA, and COFFEE are taught), ensuring a manageable amount of content without overwhelming the user.

Figure 7 depicts the real-time interaction within the game. We show a participant wearing a VR headset on the right side, while the left side mirrors the actual VR view the participant is seeing. In the virtual environment, the signer's hands are visible in black, and the avatar waits for the user to sign. After three seconds, the system assesses the accuracy of the sign produced by the participant. If their sign production aligns with the correct ASL sign, the teacher moves on to the next sign in the sequence, allowing for a fluid learning experience. If the sign performed by the participant is incorrect, the teacher responds by giving feedback (e.g., a head shake) and repeating the same sign, giving the signer more time to produce a correct sign. This corrective loop will be repeated up to three times if the user continues to sign incorrectly.

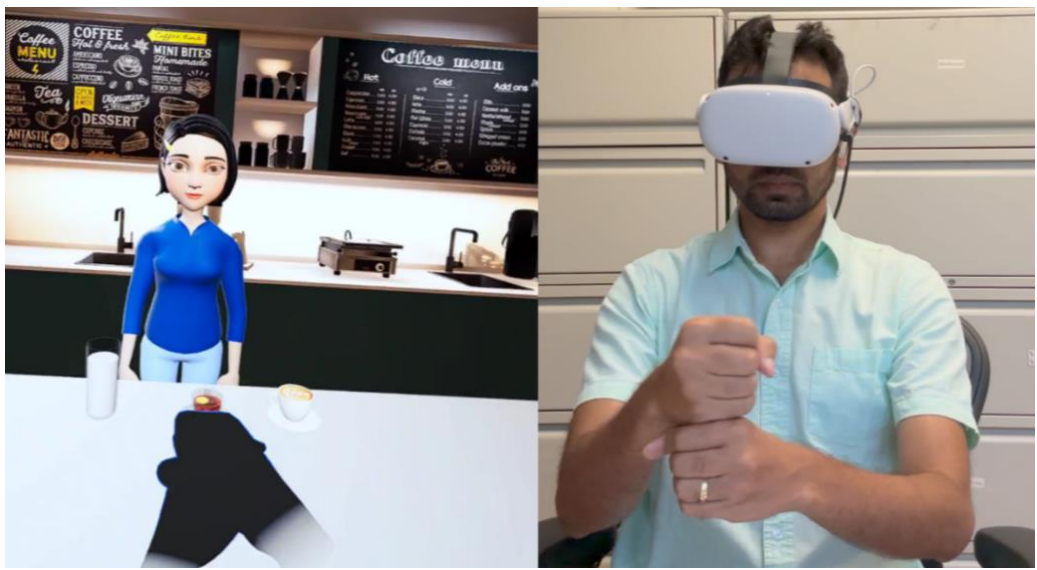

Figure 7. A side-by-side view of the system. The left and right sides show the VR and real-world views, respectively. The user is signing COFFEE, and the avatar is waiting to provide feedback.

The recognition system is highly sensitive to rotation and orientation; even a sign with similar hand and finger movements but is oriented differently will not be detected as the correct sign. During one of our test sessions, a participant signed the word "COFFEE" in a clockwise hand motion rather than a counter-clockwise hand motion, resulting in the system detecting it as an incorrect sign. Initially, this discrepancy was confusing, but upon further examination, it became evident that despite the visual similarity of the signs, their rotational hand direction was wrong, leading to the system flagging it as a wrong sign. This level of nuance is important in ASL recognition, as even the slightest variation in hand and finger movement can convey a different meaning. While learning these details can be difficult for novices, attention to these details will lead to developing precise signing skills.

## 6.1 Evaluation of the deep learning model

In this section, our primary focus will be presenting quantitative results from testing the sign recognition model. While qualitative results are valuable in real-life scenarios, emphasizing quantitative outcomes is essential for facilitating technical implementation and gaining a deeper understanding of the AI model. To refine the network's parameters, we integrated a Keras tuner into our workflow. The model is subjected to training comprising 1000 epochs, with each epoch processing a batch size of 512 samples. This tuning process has generated reasonable results, with training accuracy reaching 90.12%, demonstrating the network's capacity to learn and generalize effectively on training data. The validation accuracy, an important parameter for evaluating the model's ability to function on unknown data, was 89.37%. We tested the model's performance on a different dataset of 180 samples and then noted whether it was correctly classified. The test accuracy, an important predictor of the model's real-world performance, was 86.66%. These findings highlight the network's robustness and ability to generate consistent and accurate predictions over various datasets and circumstances.

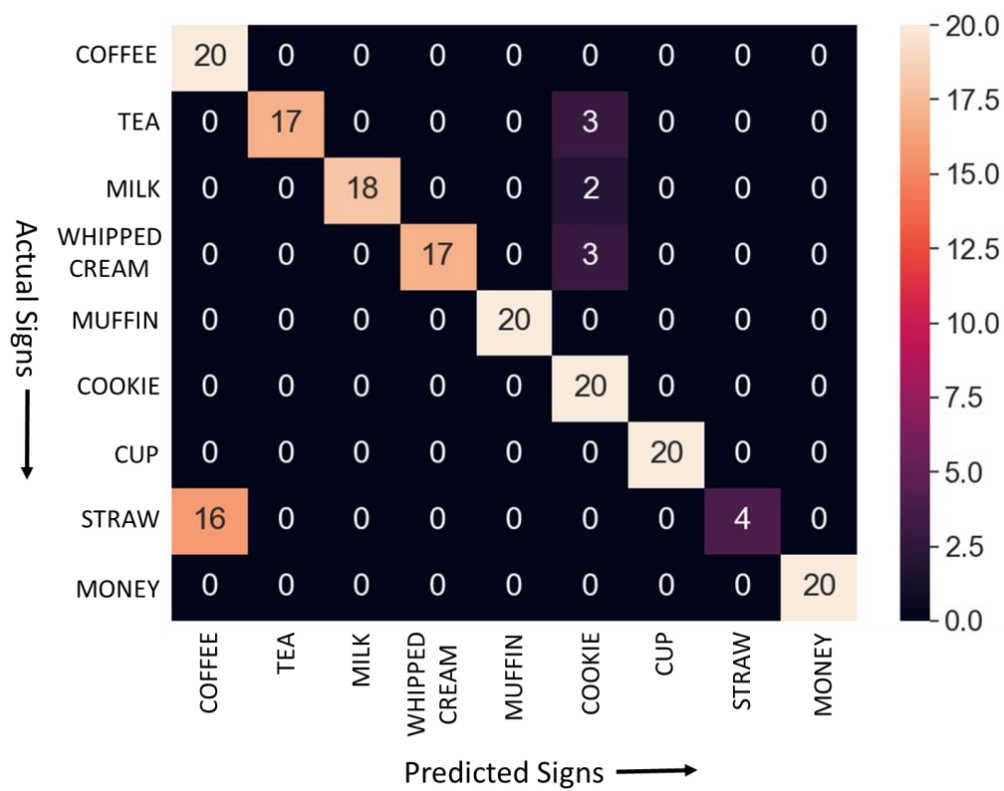

Figure 8. Confusion matrix illustrating the accuracy of sign recognition along with its corresponding frequency. The diagonal axis represents the actual accuracy of sign recognition.

To gain deeper insights into our data representation and assess the accuracy of our model's predictions, we constructed a confusion matrix, as shown in Figure 8. The shade of the color shows the frequency of the recognition. High frequencies along the diagonal axis represent higher recognition accuracies. Each digit represents the number of times the produced sign on the vertical axis was recognized as any sign on the horizontal axis. For instance, the signs COFFEE, MUFFIN, COOKIE, CUP, and MONEY all achieved 100% accuracy, highlighting the model's proficiency in correctly identifying these signs. In contrast, when the user signed STRAW, it was often erroneously classified as COFFEE, constituting a significant recognition error. The TEA and WHIPPED CREAM signs were mistakenly detected three times as COOKIE. The observed disparity can be ascribed to the resemblance in left-hand shapes employed in these signs, all necessitating the use of two hands, where one hand remains stationary while the other moves. These common misclassifications underscore the need for a more advanced algorithm to enhance the model's performance. Our ongoing efforts aim to mitigate these issues and boost the accuracy of our sign recognition system.

## 6.2 **User Experiences**

After completing the working prototype, we recruited hearing participants to evaluate the overall ASL Champ! game experience. All procedures were approved by the relevant Institutional Review Board (IRB-XXXX-XXX) in 2023 in alignment with the Declaration of Helsinki. Data related to this user experience study is available at X. All participants provided informed consent and were compensated for their time. We gathered evaluations from twelve participants, of whom three were men and 9 were women (average age = 36.6, SD = 10.3). All participants were new to learning sign language, with at most a basic understanding of ASL consisting of knowing numbers and letters. Five participants reported having never used a VR device, four reported they had used VR "once or a few times," and three have used VR "many times."

We recorded video from two angles (front-facing and side-facing) for the duration of the session. First, participants completed a demographic questionnaire (age, sex, education level, and language use). They also completed a questionnaire asking for their experience with signed languages, familiarity with signing avatars, VR, virtual assistants, and technology use. We asked about their attitudes toward technology.

Next, participants were instructed to use a "concurrent think-aloud" approach to using the ASL Champ! game (Charters, 2003). Participants were told the basics of what to expect in the ASL Champ! game and then fitted with the VR headset. The game started, and participants freely narrated their experiences. Each participant completed the prototype round of the game, in which they learned ASL signs for MILK, TEA, and COFFEE, via the interactive learning exchange described above. The approximate time participants engaged in the game with the concurrent think-aloud technique was 5-10 minutes. After removing the VR headset, we then asked them to fill out a quantitative questionnaire in which they rated eleven items from "strongly disagree" to "strongly agree" (see Table 1).

Table 1: Comprehensive list of user experience study questions

| Variable name | Question text | Response type |
|---|---|---|
| ASLexp | How much experience do you have with signed languages? | Very basic, basic, communicative, intermediate, expert |
| ASLyrs | How many years of experience do you have with signed languages? | 0, 0-1, 1-2, 2-3, 3+ |
| AvatarSeen | Before today, have you ever seen a signing avatar? | Yes/No |
| VRuse | Have you used an immersive virtual reality headset (e.g., Oculus Quest)? | Never, Once or a few times, Many times |
| VirtAssist | How much experience do you have using "virtual assistants" (e.g., Alexa, Siri)? | None at all, A little, A moderate amount, A lot, A great deal |
| TechComfort | In general, how comfortable are you with new kinds of technology? | Very uncomfortable, Slightly uncomfortable, Neither comfortable or uncomfortable, Slightly comfortable, Very comfortable |
| SelfExplan | My experience entering ASL Champ for the first time was self-explanatory. | 5-point Likert scale, Strongly Disagree, Somewhat Disagree, Neither, Somewhat |
| Intuitive | My experience navigating the interface and clicking buttons was intuitive. | |

| AppPleasant | The avatar's appearance looked pleasant to me. | Agree, Strongly Agree |
|---|---|---|
| SignRecog | My signs were accurately recognized in the game. | |
| NatMove | The avatar's movements looked natural to me. | |
| InstrExplan | The game instructions and prompts were self-explanatory. | |
| Familiar | The game elements looked familiar to me. I feel that I have seen similar elements in my previous experiences interacting with technology. | |
| UndAvatar | It was easy to understand the avatar's signs while interacting with them. | |
| AvMovement | The avatars' movements were easy for me to follow. | |
| IntUnderstand | The game interface components were sufficient for me to understand the whole process of navigating the game as a user. | |
| Immersed | I felt that I was fully and adequately immersed into the game. In other words, it felt like I was really in the coffee shop space. | |

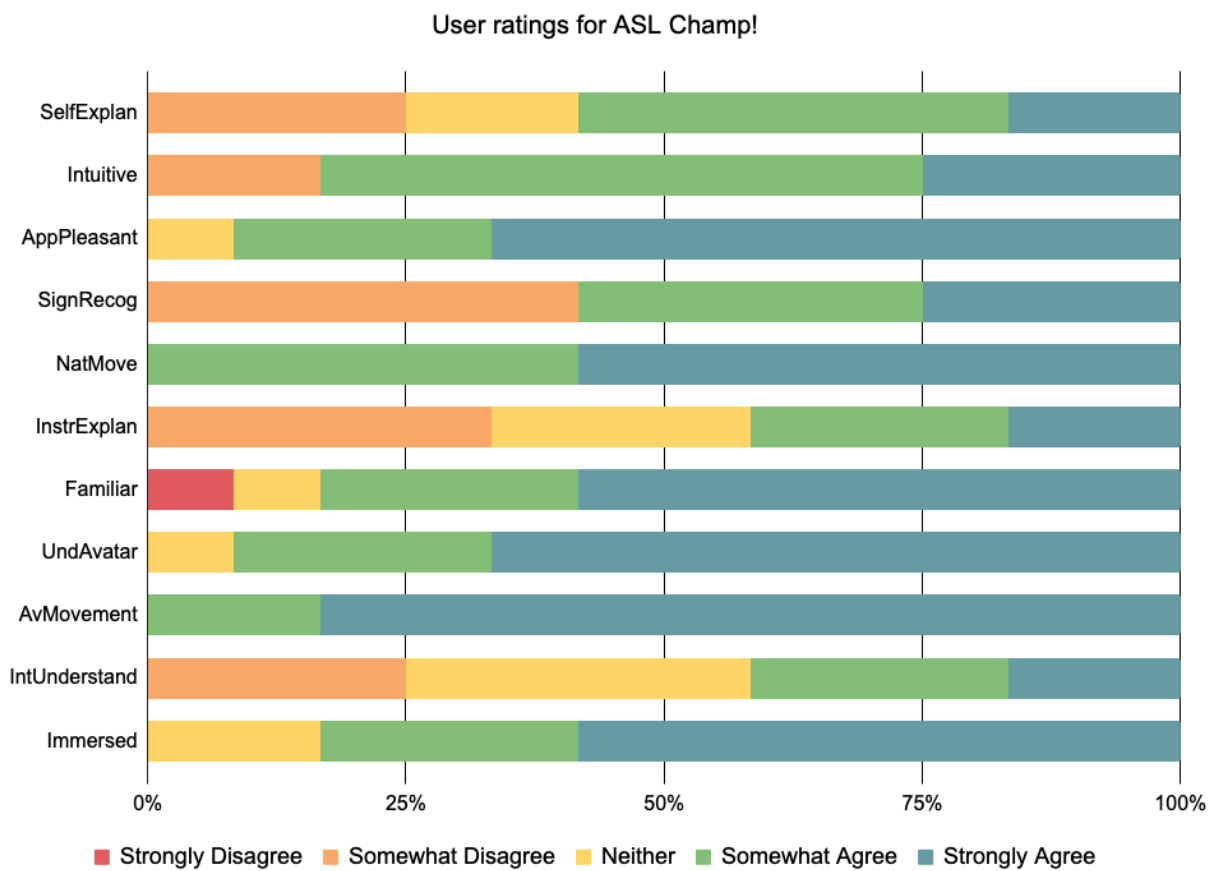

Figure 9. Descriptive results showing how participants rated eleven aspects of the ASL Champ! game after interacting with the game.

Descriptive statistics (Figure 9) show that participants found the avatar to have a pleasant experience (AppPleasant) and were generally responding well to her movements (NatMove; AvMovement). The areas we note for improvement in future versions of ASL Champ! lie primarily in the ease of navigating the interface (SelfExplan; IntUnderstand; InstrExplan) and in the success of the sign recognition model (SignRecog).

In examining the correlations between participants' backgrounds in ASL, VR, and their perceptions of a user interface, some notable findings emerged. ASL experience (ASLexp) showed a moderate positive correlation with understanding the avatar (UndAvatar; r = .53, p = .077). Participants with more years of ASL (ASLyrs) found the appearance less pleasant (AppPleasant;

r = -.42, p = .175) and the interface harder to understand (IntUnderstand; r = -.42, p = .176). Those who had seen avatars before (AvatarSeen) understood them better in this context (r = .67, p = .016). VR use (VRuse) correlated positively with finding the interface intuitive (r = .42, p = .175). These correlations, while insightful, are not all statistically significant at the p < .05 level, highlighting the need for cautious interpretation.

While participants generally felt comfortable interacting with the coffee shop environment and the avatar, some expressed confusion about the task requirements. One participant attempted to grab objects instead of mimicking the sign, indicating a need for clearer instructions. Additionally, the timing constraints of the sign recognition system posed a challenge for some participants. They had to press the space button and produce the sign within three seconds to receive feedback. However, some participants initiated the action earlier or later, leading to incorrect sign recognition. To address these issues, future system development should incorporate more comprehensive instructions and offer optional product tours to ensure participants fully grasp the task objectives and execution methods.

## 7 CONCLUSION

In this work, we developed a virtual ASL learning program with real-time avatar-driven feedback, and presented a user experience study. Our ASL sign data collection process involved diverse ASL users, encompassing variations in hand shapes, spatial positioning, orientation, and movement patterns. This approach closely resembles the unpredictability of real-world sign language interactions, where constraints are minimal and variability is abundant. The ASL Champ! prototype features dual hand-based dynamic signs with acceptable recognition accuracy. While a simple coffee shop sign may not suffice for implementing a complete learning system, it represents a small step toward a significant milestone. Our efforts created a powerful deep-learning model that was fine-tuned for precise sign detection.

It is essential to acknowledge that our sign recognition training dataset was limited to 15 signers; in our future work, we will record sign data from more people to gather a more widely representative sample for the sign recognition model. The confusion matrix reveals an unexpectedly low accuracy in one of the sign recognition categories. In response, we are committed to developing a more robust deep-learning model to elevate the accuracy and overall performance of sign recognition. Additionally, we have not yet incorporated non-manual markers (e.g., eyebrow or mouth movements) into the current version of *ASL Champ!*. Recognizing this, all signs we included are easily understood without non-manual marking. Consequently, the avatar does not express facial expressions and eye gaze and does not recognize them from the user. Some participants in our user study expressed a desire for the avatar to mirror their eye direction. Our future development roadmap includes the integration of non-manual markers, encompassing elements like facial expressions, eye movements, and body gestures to enrich the user experience. Overall, the work presented here describes a working prototype of an ASL-learning game in virtual reality, including a critical component in which the system recognizes and corrects users' signing attempts. While there is more work to be done to achieve a fully-fledged ASL learning game in VR, this interactive learning scenario promises an engaging and semi-realistic manner of learning a new signed language.

Acknowledgments

The authors are grateful to {name} and {name} for their assistance with the project. We gratefully acknowledge support from the {funding body} under grant #XXXXXXX.